\documentclass[aps,prx,superscriptaddress,reprint]{revtex4-2}
\usepackage{amsmath}
\usepackage{amssymb}
\usepackage{graphicx}
\usepackage{url}
\usepackage{hyperref}
\usepackage{color,xcolor}
\usepackage{epstopdf}
\usepackage{float}
\usepackage{ulem}
\usepackage[percent]{overpic}
\usepackage{siunitx}

\begin{document}

\title{Block Mott insulating state induced by next-nearest neighbor hopping in the $S = 3/2$ zigzag chain BaCoTe$_2$O$_7$}
\author{Ling-Fang Lin}
\author{Yang Zhang}
\affiliation{Department of Physics and Astronomy, University of Tennessee, Knoxville, Tennessee 37996, USA}
\author{Gonzalo Alvarez}
\affiliation{Computational Sciences and Engineering Division, Oak Ridge National Laboratory, Oak Ridge, TN 37831, USA}
\author{Adriana Moreo}
\author{Elbio Dagotto}
\affiliation{Department of Physics and Astronomy, University of Tennessee, Knoxville, Tennessee 37996, USA}
\affiliation{Materials Science and Technology Division, Oak Ridge National Laboratory, Oak Ridge, Tennessee 37831, USA}

\begin{abstract}
Quasi-one-dimensional correlated electronic multi-orbital systems with either ladder or chain geometries continue attracting considerable interest due to their complex electronic phases arising from the interplay of the hopping matrix, the crystal-fields splitting, the electronic correlations (Hubbard repulsion $U$ and Hund coupling $J_{\rm H}$), and strong quantum fluctuations. Recently, the intriguing cobalt zigzag chain system BaCoTe$_2$O$_7$, with electronic density $n = 7$, was prepared experimentally. Here, we systematically study the electronic and magnetic properties of this quasi-one-dimensional compound from the theory perspective. Based on first-principles density functional theory calculations, strongly anisotropic one-dimensional electronic Co $3d$ bands were found near the Fermi level. By evaluating the relevant hopping amplitudes, we provide the magnitude and origin of the nearest-neighbor (NN) and next nearest-neighbor (NNN) hopping matrices in BaCoTe$_2$O$_7$. With this information, we constructed a three-orbital electronic Hubbard model for this zigzag chain system, and studied two cases: with only a NN hopping matrix, and with NN plus NNN hopping matrices. Introducing the Hubbard and Hund couplings and studying the model via the density matrix renormalization group method, we constructed the ground-state phase diagram. A robust staggered $\uparrow$-$\downarrow$-$\uparrow$-$\downarrow$ antiferromagnetic (AFM) region was found when only the NN hopping matrix in the chain direction was employed. However,
for the {\it realistic} case where the NNN hopping matrix is also included, the dominant state becomes instead a
block AFM $\uparrow$-$\uparrow$-$\downarrow$-$\downarrow$ order, in agreement with experiments. The system displays Mott insulator characteristics with three half-filled orbitals, when the block AFM order is stable. Our results for BaCoTe$_2$O$_7$ provide guidance to experimentalists and theorists working on this zigzag one-dimensional chain and related materials.
\end{abstract}

\maketitle

\section{Introduction}
Because of their intertwining charge, spin, and lattice degrees of freedom as well as strong quantum fluctuations~\cite{Dagotto:rmp94,Grioni:JPCM,Monceau:ap,Dagotto:Rmp}, a variety of fascinating physical properties have been reported in one-dimensional (1D) correlated electronic systems, such as high-critical temperature superconductivity~\cite{cu-ladder1,cu-ladder2,cu-ladder3,Takahashi:Nm,Ying:prb17,Zhang:prb17,Zhang:prb18}, and charge density waves~\cite{Zhang:prb21,Zhang:prbcdw,Gooth:nature}, to name a few.

Further more, when the 1D system contains several active orbitals, further intriguing properties have been unveiled arising from the interplay among the hopping matrix, the crystal-field splittings, and electronic correlations where in addition to the canonical Hubbard repulsion $U$, also the Hund coupling $J_{\rm H}$ plays a key role. For example, considering the competition between hopping and electronic correlations in the intermediate coupling range region, the exotic orbital-selective Peierls phase~\cite{Streltsovt:prb14} and orbital-selective Mott phase~\cite{osmp}, with a mixture of localized and itinerant behavior of the different orbitals, were obtained for some real 1D systems~\cite{Zhang:ossp,Patel:osmp,Herbrych:osmp}. Furthermore, a large interorbital electronic hopping could lead to a ferromagnetic (FM) insulating state between doubly occupied and half-filled orbitals~\cite{Lin:prl21,Lin:cp}, which potentially is already realized in some iron chain materials~\cite{McCabe:cc,McCabe:prb,Lin:prb,NaFe2X2}. Varying the electronic densities and electronic correlations, many complex and interesting spin orders were obtained by the competition between FM vs AFM tendencies~\cite{Zhang:prbblock,Herbrych:prbblock}.

\begin{figure}
\centering
\includegraphics[width=0.48\textwidth]{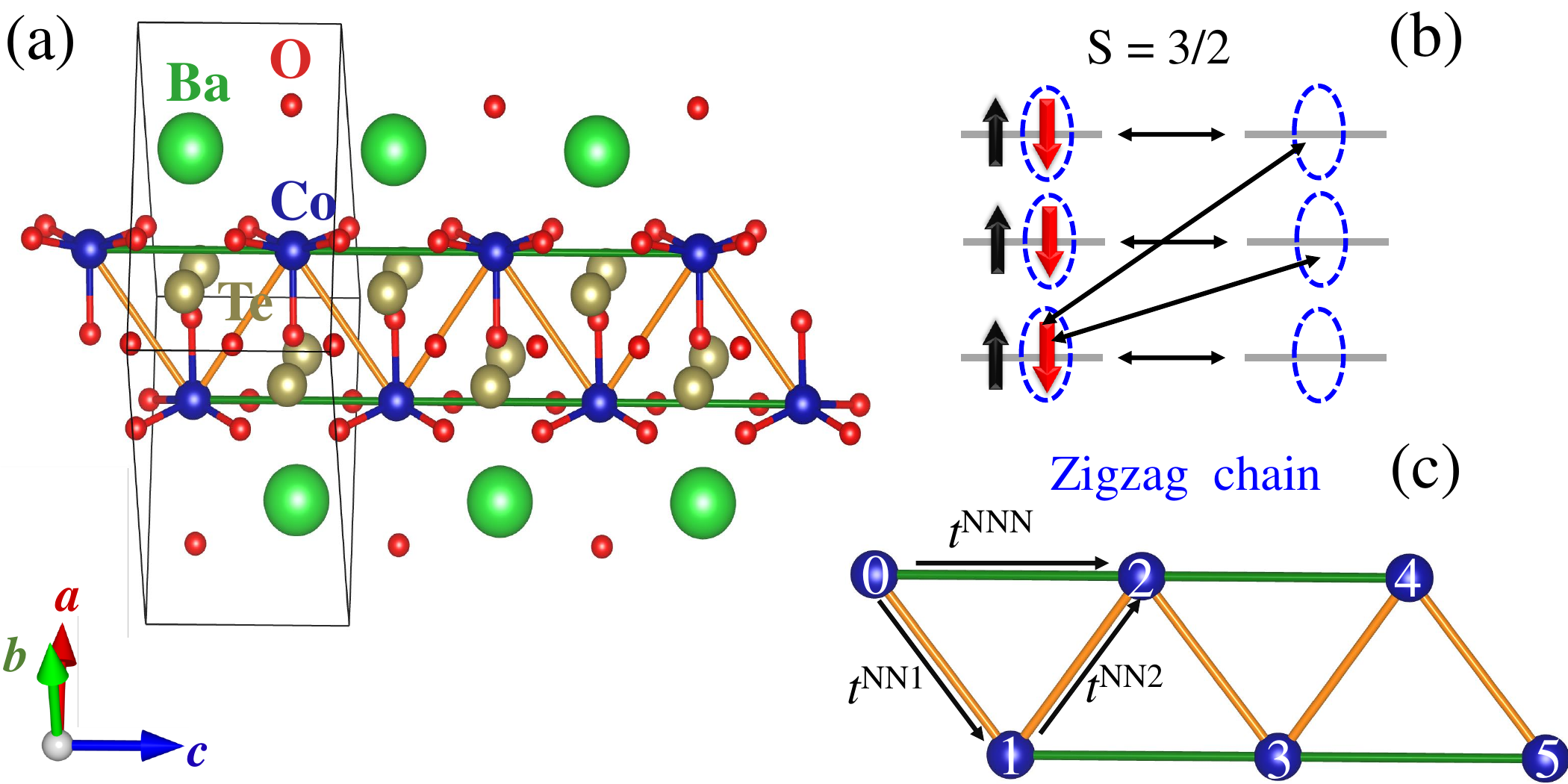}
\caption{(a) Schematic crystal structure of the BaCoTe$_2$O$_7$ conventional cell (green = Ba; blue = Co; brown = Te; red = O.). (b) AFM superexchange path for two NN sites caused by both intraorbital and interorbital hoppings with $S = 3/2$.
(c) Schematic lattice of zigzag chain.}
\label{Zigzag}
\end{figure}

Recently, a cobalt-based zigzag chain compound BaCoTe$_2$O$_7$ has been systematically studied using neutron diffraction experiments. An interesting ``block'' AFM state with a $\uparrow$-$\uparrow$-$\downarrow$-$\downarrow$ pattern was found along the zigzag chain direction~\cite{Li:scpma}. BaCoTe$_2$O$_7$ has an orthorhombic structure with space group $Ama2$ (No. 40), as shown in Fig.~\ref{Zigzag}(a), where the nearest-neighbor (NN) Co ions are connected by alternating inverted square-pyramides CoO$_5$. A Co$^{2+}$ ion with the $d^7$ configuration has three half-filled and two double-occupied orbitals, leading to a net $S = 3/2$ state. In this case, due to Pauli's principle, both interorbital and intraorbital hoppings would lead to AFM coupling between two Co sites, as displayed in Fig.~\ref{Zigzag}(b). However, compared with the straight uniform chain, in the zigzag chain the next nearest-neighbor (NNN) hopping will be enhanced due to the reduced distance of the NNN bonds (see Fig.~\ref{Zigzag}(c)). In BaCoTe$_2$O$_7$, the NNN Co-Co bond is about $\sim 5.574$ \AA, which is close to that of the NN Co-Co bond ($\sim 4.658$ \AA). As a result, the NNN hopping can be comparable to the NN hopping, leading to strong AFM coupling both in the NN and NNN bonds, resulting in a strong magnetic frustration. What kind of spin state will dominante in this environment?

BaCoTe$_2$O$_7$ belongs to a noncentrosymmetric polar material family Ba$M$Te$_2$O$_7$ ($M$ = Mg, Co, Ni, Cu and Zn)~\cite{Yeon:ic1,Yeon:ic2,Chen:prm}. BaMgTe$_2$O$_7$ and BaZnTe$_2$O$_7$ are nonmagnetic~\cite{Yeon:ic2}. Moreover, no long-range magnetic ordering was found down to 1.8 K in BaCuTe$_2$O$_7$ but with a broad peak around 71 K in the magnetic susceptibility~\cite{Yeon:ic1}. BaNiTe$_2$O$_7$ has a commensurate AFM structure (0.5, 1, 0), also involving the $\uparrow$-$\uparrow$-$\downarrow$-$\downarrow$ coupling along the chain direction, as in the case of Co. Although there are many experimental studies in this family of materials, systematic theoretical studies are still rare.

To better understand the electronic and magnetic properties, here both first-principles density functional theory (DFT) and density matrix renormalization group (DMRG) methods were employed to investigate BaCoTe$_2$O$_7$. First, the {\it ab initio} DFT calculations indicate a strongly anisotropic electronic structure for BaCoTe$_2$O$_7$, in agreement with its anticipated 1D zigzag geometry.  Based on the Wannier functions obtained from first-principle calculations, we obtained the relevant hopping amplitudes and on-site energies of the cobalt atoms. For the NN hopping matrix, the largest hopping arises from the $d_{3z^2-r^2}$ orbital. Intriguingly, for the NNN hopping matrix, the largest hopping element emerges from $d_{xy}$ to $d_{xy}$, and microscopically this is caused by the super-super exchange via a complex path $d_{xy}$-$p_x$/$p_y$-$p_x$/$p_y$-$d_{xy}$. Anticipating rich results, we constructed a multi-orbital Hubbard model for the cobalt zigzag chains considering both NN and NNN hoppings.

Based on DMRG calculations, we obtained the ground-state phase diagram varying the on-site Hubbard repulsion $U$ and the on-site Hund coupling $J_{\rm H}$. When the NNN hoppings are properly included, the block AFM $\uparrow$-$\uparrow$-$\downarrow$-$\downarrow$ state with Mott insulating (MI) characteristics was found to be dominant in a robust portion of the phase diagram, in agreement with the experimental results. In addition, paramagnetism was found in the regime of weak Hubbard coupling strength. Using DFT+$U$, the block spin order was here also found to be the most likely magnetic ground state compared to other magnetic orders, in agreement with experiments.
Then, both techniques used here agree that a block arrangement is the most stable for this compound. Note that in Ref.~~\cite{Li:scpma}, where the experimental result for the block phase was reported, the theoretical component also used DFT+$U$ for the block state
but without comparing with other possible states. Thus, our effort here
for the first time reports that the block phase is indeed the ground state from
a microscopic perspective using two independent techniques.

\section{Methods}

\subsection{DFT Method}
In this work, we employed first-principles DFT calculations using the Vienna {\it ab initio} simulation package (VASP) software within the projector augmented wave (PAW) method~\cite{Kresse:Prb,Kresse:Prb96,Blochl:Prb}, where
the electronic correlations were considered by using the generalized gradient approximation (GGA) with the Perdew-Burke-Ernzerhof functional~\cite{Perdew:Prl}. The plane-wave cutoff used was $520$ eV
and the $k$-point mesh was $6\times6\times3$ for the
calculations of the electronic structure of the non-magnetic state, which was accordingly adapted for the magnetic calculations. To obtain the hopping matrix and crystal-field splitting parameters, the maximally localized Wannier functions (MLWFs) method was employed to fit the Co's five $3d$ bands by using the WANNIER90 packages~\cite{Mostofi:cpc}. To better understand the magnetic properties, the local spin density approach (LSDA) plus $U_{\rm eff}$ with the Dudarev format was employed~\cite{Dudarev:prb}. Both the lattice constants and atomic positions were fully relaxed with different spin configurations until the Hellman-Feynman force on each atom was smaller than $0.01$ eV/{\AA}. All the crystal structures were visualized with the VESTA code~\cite{Momma:vesta}.

\subsection{Multi-orbital Hubbard Model}
To understand the magnetic properties of the one-dimensional zigzag chain, we employed the standard
multi-orbital Hubbard model, which includes a kinetic energy component and Coulomb interaction energy terms
$H = H_k + H_{int}$. The tight-binding kinetic portion is described as:
\begin{eqnarray}
H_k = \sum_{\substack{<i,j>\sigma\gamma\gamma'}}t_{\gamma\gamma'}(c^{\dagger}_{i\sigma\gamma}c^{\phantom\dagger}_{j\sigma\gamma'}+H.c.)+ \sum_{i\gamma\sigma} \Delta_{\gamma} n_{i\gamma\sigma},
\end{eqnarray}
where the first part represents the hopping of an electron from orbital $\gamma$ at site $i$ to orbital $\gamma'$ at the NN or NNN site $j$, using a chain of length $L$. $\gamma$ and $\gamma'$ represent the three different orbitals. The second part are the crystal fields.

The standard electronic interaction portion of the Hamiltonian is:
\begin{eqnarray}
H_{int}= U\sum_{i\gamma}n_{i\uparrow \gamma} n_{i\downarrow \gamma} +(U'-\frac{J_{\rm H}}{2})\sum_{\substack{i\\\gamma < \gamma'}} n_{i \gamma} n_{i\gamma'} \nonumber \\
-2J_{\rm H}  \sum_{\substack{i\\\gamma < \gamma'}} {{\bf S}_{i\gamma}}\cdot{{\bf S}_{i\gamma'}}+J_{\rm H}  \sum_{\substack{i\\\gamma < \gamma'}} (P^{\dagger}_{i\gamma} P^{\phantom\dagger}_{i\gamma'}+H.c.).
\end{eqnarray}
The first term is the intraorbital Hubbard repulsion. The second term is the electronic repulsion between electrons at different orbitals where the standard relation $U'=U-2J_{\rm H}$ is assumed due to rotational invariance. The third term represents the Hund's coupling between electrons occupying the Co's $3d$ orbitals. The fourth term is the pair hopping between different orbitals at the same site $i$, where $P_{i\gamma}$=$c_{i \downarrow \gamma} c_{i \uparrow \gamma}$.

To solve the multi-orbital Hubbard model, by introducing quantum fluctuations, the many-body technique that we employed was based on the DMRG method~\cite{white:prl,white:prb}, where specifically we used the DMRG++ software package~\cite{Alvarez:cpc}. In our DMRG calculations, we employed a $16$-sites cluster chain with three orbitals per site and open-boundary conditions (OBC). Furthermore, at least $1200$ states were kept and up to $21$ sweeps were performed during our DMRG calculations. In addition, the average electronic filling $n = 3$ for the three orbital at each site was considered.

In the tight-binding term, we used the Wannier function basis \{$d_{3z^2-r^2}$, $d_{yz}$,$d_{xy}$\}, here referred to as $\gamma$ =  \{0, 1, 2\}, respectively. We only considered the NN and NNN hopping matrix:
\begin{equation}
\begin{split}
t^{NN1}_{\gamma\gamma'} =
\begin{bmatrix}
    -0.079	&       0.027	  &  0.028	\\ 
     0.027	 &   0.022	   &  0.009	\\
    -0.028	&     -0.009	  &     -0.003	
\end{bmatrix}.\\
\end{split}
\end{equation}
\begin{equation}
\begin{split}
t^{NN2}_{\gamma\gamma'} =
\begin{bmatrix}
    -0.079	 &   -0.027	    &  0.028	\\
    -0.027	  &   0.022	   &   -0.009	\\
    -0.028	   & 0.009	   & -0.003	
\end{bmatrix}.\\
\end{split}
\end{equation}
\begin{equation}
\begin{split}
t^{NNN}_{\gamma\gamma'} =
\begin{bmatrix}
    -0.026	&     -0.007	    &  0.019	\\
     0.007	&    0.013	    &-0.038	\\
    -0.019	&     -0.038	   & 0.124	
\end{bmatrix}.\\
\end{split}
\end{equation}

All the hopping matrix elements are given in eV units. $\Delta_{\gamma}$ is the crystal-field splitting of orbital $\gamma$. Specifically, $\Delta_{0} =-0.072$, $\Delta_{1} = -0.397$, and $\Delta_{2} =0.477$ (the Fermi level is considered to be zero). Note that in the notation convention we used, as shown in Fig.~\ref{Zigzag}(c), the hopping matrices have direction. For example, the hopping matrix from atom 0 to atom 1 is $t^{NN1}$ and the one from atom 1 to atom 0 is the transposed of $t^{NN1}$.

\section{Results}

\subsection{Crystal-field splitting and the origin of strong NNN hopping}
\begin{figure}
\centering
\includegraphics[width=0.48\textwidth]{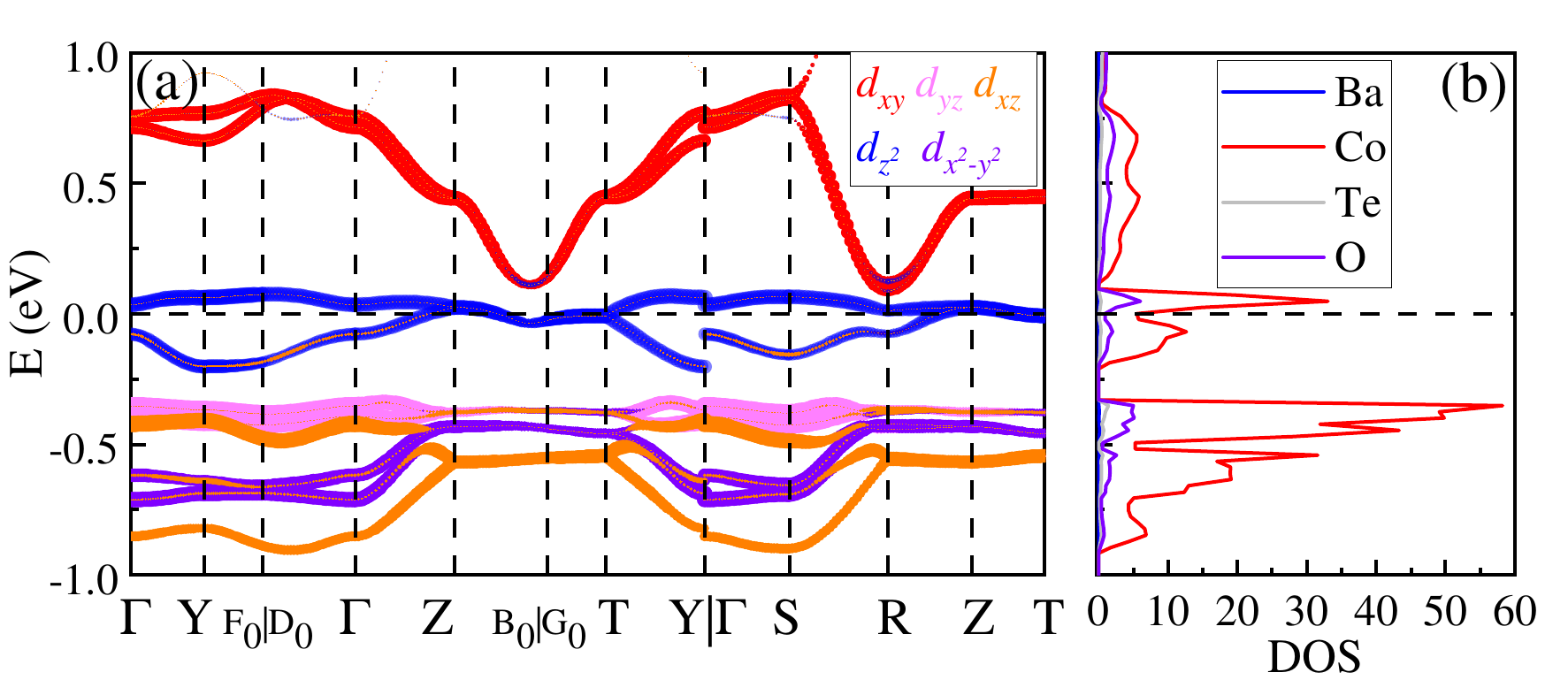}
\caption{(a) Projected band structure of BaCoTe$_2$O$_7$ for the non-magnetic state. The Fermi level is shown with a dashed horizontal line. The weight of each Co's $3d$ orbital is given by the size of the circles. Note that the local $z$ axis is perpendicular to the CoO$_5$ plane towards the top O atom, while the local $y$-axis is along the $c$-axis, leading to the $xy$ orbital lying along the in-plane CoO bond directions. The coordinates of the high-symmetry points in the Brillouin zone are $\Gamma$ = (0, 0, 0), Y = (0.5, 0.5, 0), ${\rm F}_0$ = (0.30769, 0.69231, 0), ${\rm D}_0$ = (-0.30769, 0.30769, 0), Z = (0, 0, 0.5), ${\rm B}_0$ = (-0.30769, 0.30769, 0.5), ${\rm G}_0$ = (0.30769, 0.69231, 0.5), T = (0.5, 0.5, 0.5), S = (0, 0.5, 0), and R = (0, 0.5, 0.5). Note that all the high-symmetry points are in scaled units, corresponding to the units of $2\pi$/s, ($s = a, ~b$ or $c$). (b) Density-of-states near the Fermi level of BaCoTe$_2$O$_7$  for the non-magnetic phase (blue = Ba; red = Co; gray = Te; purple = O.). Note that the DFT electronic structures are calculated using the experimental crystal-structure information~\cite{Li:scpma}, without additional Hubbard $U$.}
\label{dosband_nm}
\end{figure}

First, we calculated the electronic structures of the non-magnetic state of BaCoTe$_2$O$_7$, as shown in Fig.~\ref{dosband_nm}, using the experimental crystal structure~\cite{Li:scpma}. As displayed in Fig.~\ref{dosband_nm}(b), near the Fermi level, the electronic density is mainly contributed by the cobalt $3d$ orbitals, slightly hybridized with O's $2p$ orbitals, where most of these O's $2p$ orbitals are located in the lower energy region (not shown here). Furthermore, the Co $3d$ sates are located in a relatively narrow range of energy from $\sim -1$ to $\sim 1$ eV, indicating a large charge-transfer energy between Co $3d$ and O $2p$ states, leading to a Mott-Hubbard system.

In addition, the band structure of BaCoTe$_2$O$_7$ clearly shows that the bands are more dispersive along the chains than along other directions, such as $\Gamma$ to Z and S to R, which is compatible with the dominant presence of 1D zigzag chains along the $c$-axis. Furthermore, the $d_{xy}$ orbital bands are more dispersive than other orbitals, indicating that $d_{xy}$ should play the primary role in magnetism and other physical properties of BaCoTe$_2$O$_7$, as displayed in Fig.~\ref{dosband_nm}(a).

\begin{figure}
\centering
\includegraphics[width=0.48\textwidth]{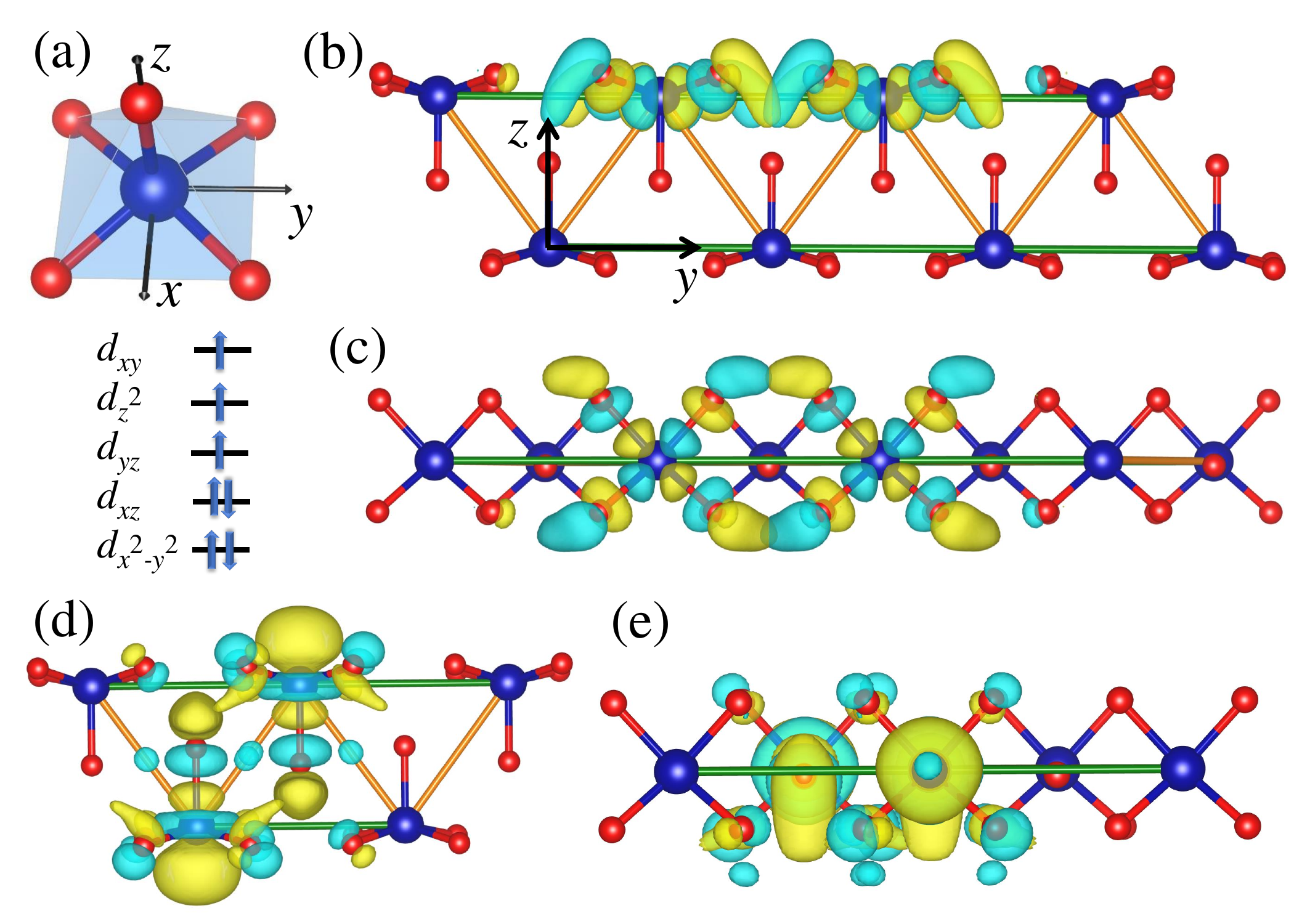}
\caption{ (a) Sketch of the CoO$_5$ cluster and the crystal splitting of the five $d$ orbitals. The orbital filling sketch is considered in the large $U$ and $J_{\rm H}$ limit. (b) Side view and (c) top view of Wannier functions for the Co $d_{xy}$ orbital corresponding to the NNN sites of BaCoTe$_2$O$_7$ (yellow and light blue indicate the two signs of the wave function). (d) Side view and (e) top view of Wannier functions for the Co $d_{3z^2-r^2}$ orbital for the NN sites of BaCoTe$_2$O$_7$.}
\label{crystal}
\end{figure}

Based on the MLWFs method, we obtained the crystal-field splittings for Co $3d$ orbitals (see Fig.~\ref{crystal}(a)) by using the WANNIER90 packages~\cite{Mostofi:cpc}. By introducing the electronic correlations and considering the high-spin state, the $d_{xz}$ and $d_{x^2-y^2}$ orbitals are fully occupied while $d_{xy}$, $d_{3z^2-r^2}$, and $d_{yz}$ are only half-filled due to the $d^7$ configurations as well, then the system will be in a $S= 3/2$ state in the large $U$ and $J_{\rm H}$ limit, as displayed in Fig.~\ref{crystal}(a). For the NN sites, the largest hopping element ($\sim 0.079$ eV) arises from $d_{3z^2-r^2}$ orbitals while the hopping between $d_{xy}$ orbitals is quite small ($\sim 0.003$ eV). However, the NNN hopping between $d_{xy}$ orbitals, $\sim 0.124$ eV, is much larger than the other intraorbital and interorbital hoppings.

To better understand these hoppings, we first plot the Wannier functions of $d_{xy}$ in Figs.~\ref{crystal}(b) and (c), where it clearly shows that the $d_{xy}$ orbital displays strong 1D characteristics along the $c$-axis, leading to a strong overlap between NNN Co-Co sites via the O's $p_x$ or $p_y$ orbitals, while the overlap is nearly zero among the NN sites. Thus, overall this leads to a strong AFM coupling among the NNN sites due to super-super exchange via  the $d_{xy}$-$p_x$/$p_y$-$p_x$/$p_y$-$d_{xy}$ path. However, the other two orbitals (specifically, $d_{3z^2-r^2}$ and $d_{yz}$) have smaller overlaps along the NNN bonds but contribute instead to the NN bonds because they are oriented along the $z$-axis. As shown in Figs.~\ref{crystal}(d) and (e), there are also obvious overlaps between two $d_{3z^2-r^2}$ orbitals along the NN bond via a mixture of the apical O's $p_z$ and in-plane O's $p$ orbitals. Thus, the physical properties of this system are determined by the combination of the influence of both NN and NNN hoppings.

\subsection{DMRG phase diagrams}

For 1D systems in general, quantum fluctuations are quite important at low temperatures but DFT neglects those fluctuations. Thus, we employ the many-body DMRG technique to incorporate the quantum effects due to the magnetic couplings along the zigzag chain. These quantum fluctuations are needed to fully clarify the true magnetic ground state properties. Here, we considered the previously described effective multi-orbital Hubbard model in the zigzag lattice with NN and NNN hopping matrix assuming three electrons in three orbitals per site in average, i.e. corresponding to the electronic density per site $n = 3$. It also should be noticed that the DMRG method has repeatedly proven to be a powerful technique for discussing low-dimensional interacting systems~\cite{Schollw:rmp05,Stoudenmire:ARCMP}. To understand the physical properties of the system under consideration here, we measured several observables based on the DMRG calculations.

First, we calculated the spin-spin correlation $S(r)$ and spin structure factor $S(q)$ at $U$ = 4 eV and $J_{\rm H}/U$ = 0.2 for two cases: (1) only NN hopping and (2) NN plus NNN hoppings, the latter being the most realistic for the compound
we considered. The spin-spin correlations in real space are defined as
\begin{eqnarray}
S(r)=\langle{{\bf S}_i \cdot {\bf S}_j}\rangle,
\end{eqnarray}
with $r=\left|{i-j}\right|$, and the spin structure factor is
\begin{eqnarray}
S(q)=\frac{1}{L}\sum_{r}e^{-iqr}S(r).
\end{eqnarray}

\begin{figure}
\centering
\includegraphics[width=0.48\textwidth]{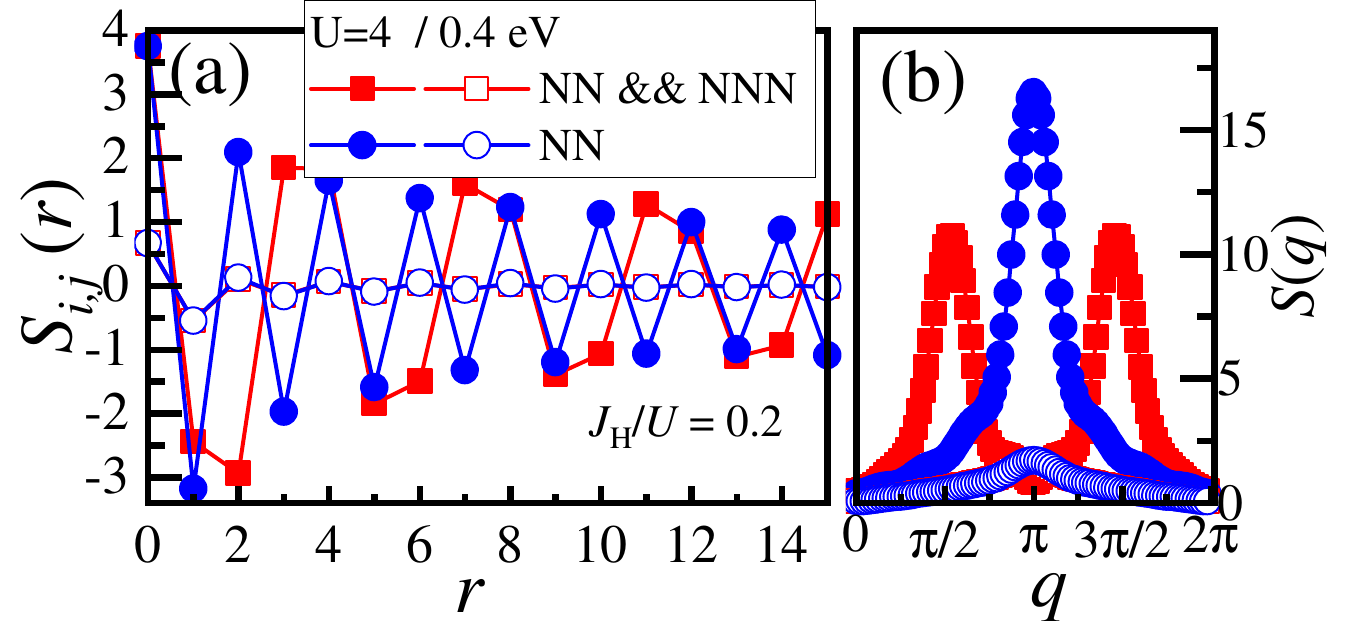}
\caption{(a) Spin-spin correlation $S(r)=\langle{{\bf S}_i \cdot {\bf S}_j}\rangle$ (with distance $r=\left|{i-j}\right|$ in real space) and (b) the spin structure factor $S(q)$, for zigzag (red line) and uniform (blue line) chains, both at $J_{\rm H}$/$U$ = 0.2 with $U = 4$~eV (solid symbols) and 0.4~eV (open symbols). We use a chain with $L = 16$.}
\label{SrSq}
\end{figure}

Figure~\ref{SrSq} shows the spin-spin correlation $S(r)$=$\langle{{\bf S}_i \cdot {\bf S}_j}\rangle$ as a function of distance $r$ at $J_{\rm H}$/$U$ = 0.2. The distance is defined as $r=\left|{i-j}\right|$, with $i$ and $j$ site indexes. For $U = 4$ eV, with only NN hopping, the spin-spin correlation $S(q)$ shows a canonical staggered $\uparrow$-$\downarrow$-$\uparrow$-$\downarrow$ AFM phase, with a peak at $\pi$ in the spin structure factor $S(q)$. However, by considering the NNN hopping, it shows a quite different spin arrangement, namely a block AFM with a $\uparrow$-$\uparrow$-$\downarrow$-$\downarrow$ pattern, corresponding to a peak at $\pi$/2 in the spin structure $S(q)$, as display in Fig.~\ref{SrSq}. Thus, these results indicate that the NNN hopping is important to understand the block AFM state in the $S = 3/2$ zigzag chain BaCoTe$_2$O$_7$. For $U = 0.4$ eV, the spin correlation $S(r)$ decays rapidly vs. distance $r$ for both cases, whether or not involving the NNN hopping matrix, due to the 1D strong quantum fluctuations and the weak value of the coupling $U$.

Next, we calculated the DMRG phase diagram for different values of  $U$ and $J_{\rm H}/U$ for the two hopping
cases mentioned above, based on the DMRG measurements of the spin-spin correlation and spin structure factor, as well as the site-average occupancy of orbitals and orbital-resolved charge fluctuations.

\begin{figure}
\centering
\includegraphics[width=0.48\textwidth]{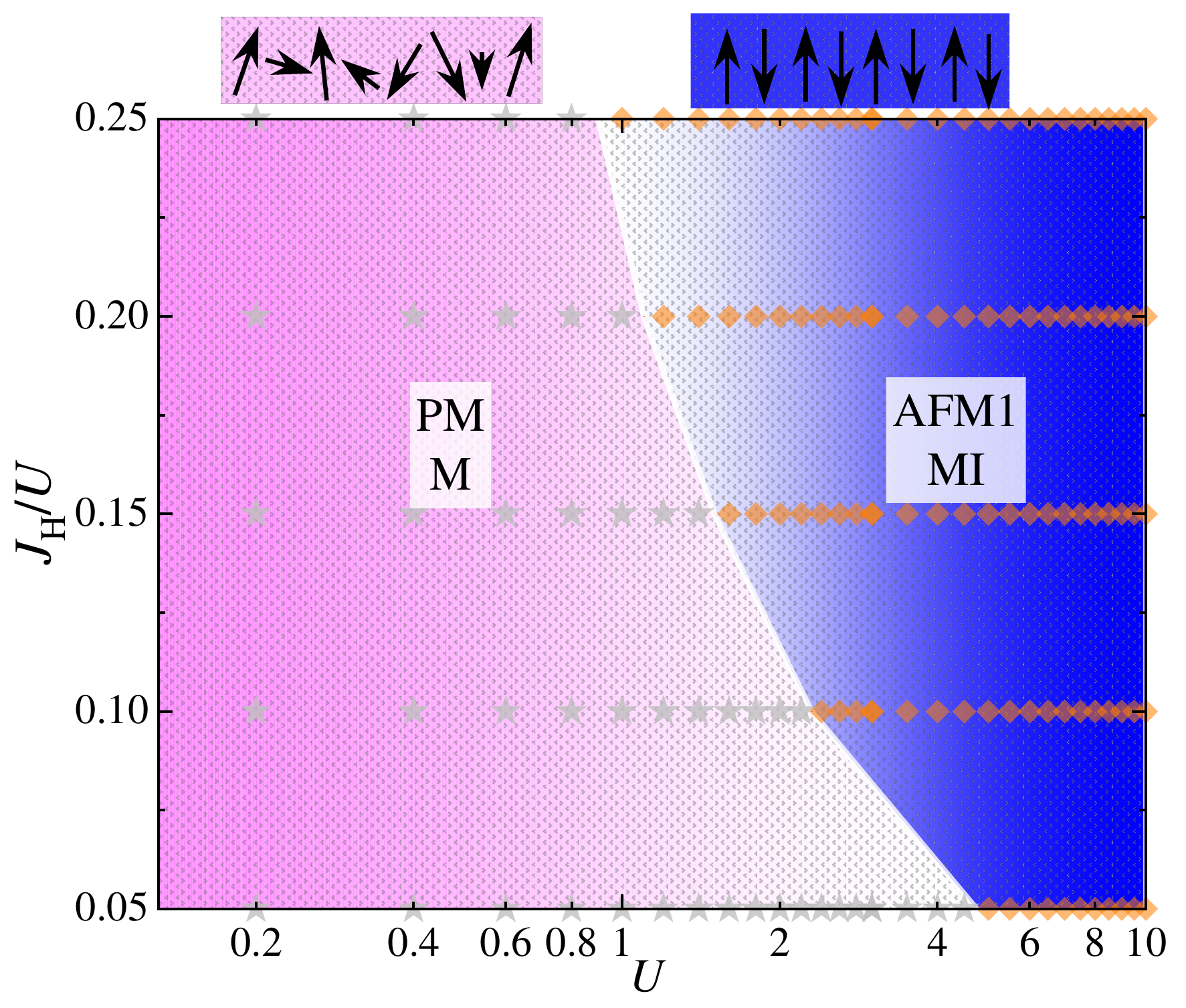}
\caption{Phase diagram of the three-orbital Hubbard model varying $U$ and $J_{\rm H}/U$, with only NN hopping by using DMRG and a $L = 16$ chain system with open boundary conditions. Different electronic and magnetic phases are indicated by solid regions and labels, including paramagnetic metal (PM M, in pink) and canonical staggered  AFM Mott insulator (AFM1 MI, in blue). Note that the boundaries should be considered only as crude approximations due to the discrete set of parameter points investigated. The phase boundaries are crudely determined based on some indicators, such as the electron density of each orbital, charge fluctuations, and the dominant peak in S(q).}
\label{phase_NN}
\end{figure}

As shown in Fig.~\ref{phase_NN}, we found a paramagnetic phase (PM) at small $U$, followed by a robust canonical staggered AFM phase with a $\uparrow$-$\downarrow$-$\uparrow$-$\downarrow$ pattern. At small Hubbard interaction, the spin correlation $S(r)$ decays rapidly vs. distance $r$, indicating paramagnetic behavior. By increasing $U$, the system turns to the canonical staggered AFM phase with the $\uparrow$-$\downarrow$-$\uparrow$-$\downarrow$ configuration in the whole region of our study. This is easy to understand since both interorbital and intraorbital hoppings would lead to AFM tendencies in between the three half-filling sites. As $J_{\rm H}/U$ increases, the critical value of $U$ for the PM-AFM1 transition decreases.

\begin{figure}
\centering
\includegraphics[width=0.48\textwidth]{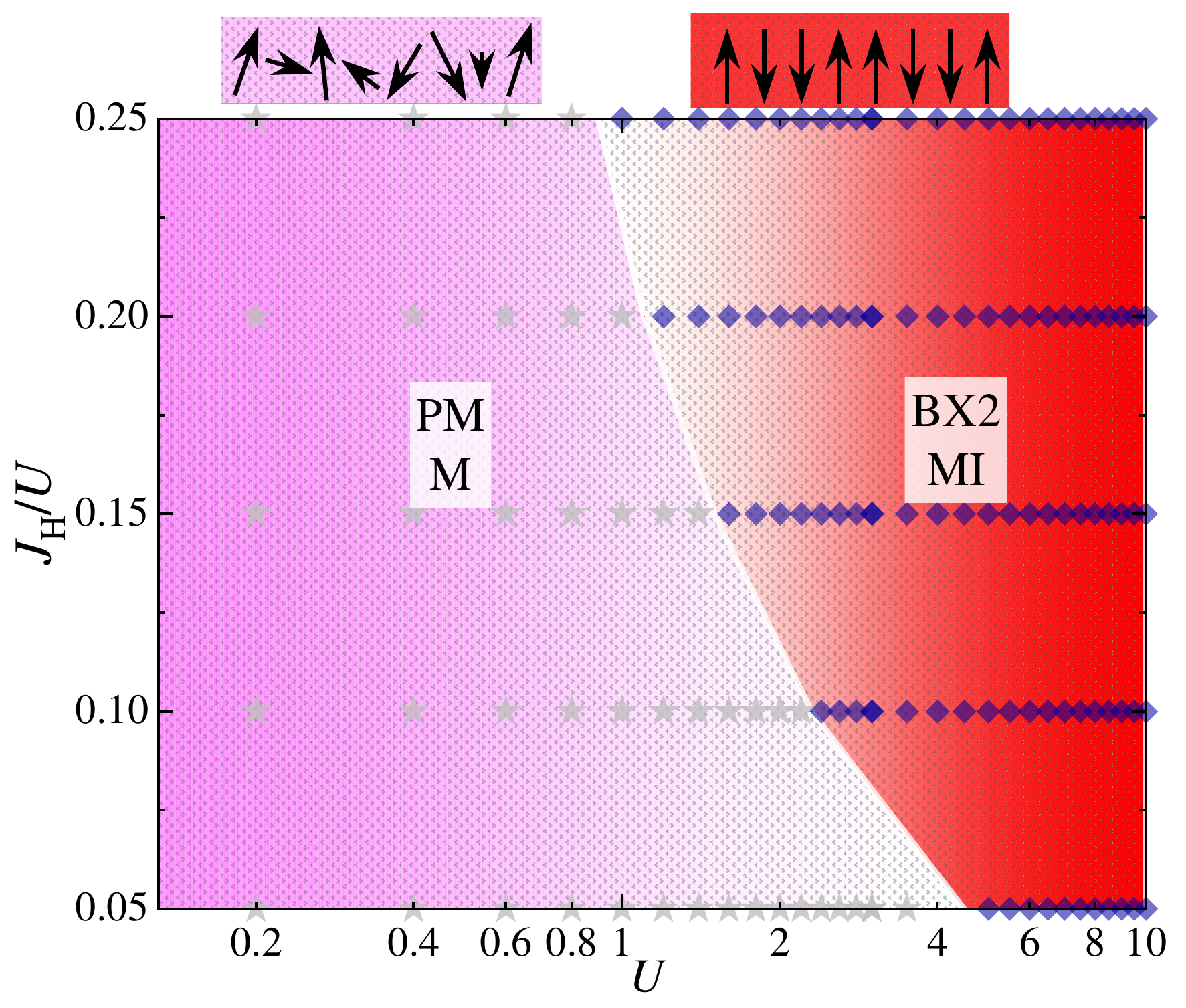}
\caption{Phase diagram of the three-orbital Hubbard model varying $U$ and $J_{\rm H}/U$, with NN plus NNN hoppings by using DMRG and an $L = 16$ chain system with open boundary conditions. Different electronic and magnetic phases are indicated by solid regions and labels, including paramagnetic metal (PM M, in pink) and Block AFM Mott insulator (BX2 MI, in red). Note that the boundaries should be considered only as crude approximations due to the discrete set of parameter points investigated. The phase boundaries are crudely determined based on some indicators, such as the electron density of each orbitals, charge fluctuations, and the dominant peak in S(q).}
\label{phase_NNN}
\end{figure}

Similarly to the case with only NN hopping, after introducing NNN hoppings the PM state was found in the small $U$ region, as displayed in Fig.~\ref{phase_NNN}. Afterwards, the Block AFM state with $\uparrow$-$\uparrow$-$\downarrow$-$\downarrow$ order is obtained by increasing $U$. Note that the BX2 state does not appear in the entire $J_{\rm H}/U$ and $U$ region explored. As $J_{\rm H}/U$ increases, the critical value of $U$ for the PM-BX2 transition is reduced, as displayed in Fig.~\ref{phase_NNN}. We do not observe any other magnetic state in the $J_{\rm H}/U$ and $U$ regions we studied. Thus, when compared to the phase diagram with only NN hopping, the NNN hopping is crucial for the stabilization of the block state in this system. This is because the intraorbital hopping of the $d_{xy}$ orbital causes the strongest AFM interaction strength to be among the NNN sites (in general following the rule that the strength is regulated by
$\sim t^2/U$) rather that among the NN sites. Thus, this system forms a block AFM pattern along the chain direction.

\subsection{PM to Block MI transition}
To understand the PM to Block phase transition and the characteristics of metallic vs insulating behavior in this system, we also studied the site-average occupancy of different orbitals $n_{\gamma}$, and orbital-resolved charge fluctuations $\delta{n_{\gamma}}$. Here, we used $J_{\rm H}/U$ = 0.2 as an example.

The site-average occupancy of orbitals, orbital-resolved charge fluctuations, and $\langle S^2_{\gamma}\rangle$ are defined as:
\begin{eqnarray}
n_{\gamma}=\frac{1}{L}\sum_{i,\sigma}{\langle}n_{i\gamma\sigma}\rangle,
\end{eqnarray}

\begin{eqnarray}
\delta{n_{\gamma}}=\frac{1}{L}\sum_{i}({\langle}n_{\gamma,i}^2\rangle-{\langle}n_{\gamma,i}{\rangle}^2),
\end{eqnarray}

\begin{eqnarray}
\langle S^2_{\gamma}\rangle =\frac{1}{L}\sum_{i}{\langle}S^2_{\gamma,i}\rangle.
\end{eqnarray}

\begin{figure}
\centering
\includegraphics[width=0.48\textwidth]{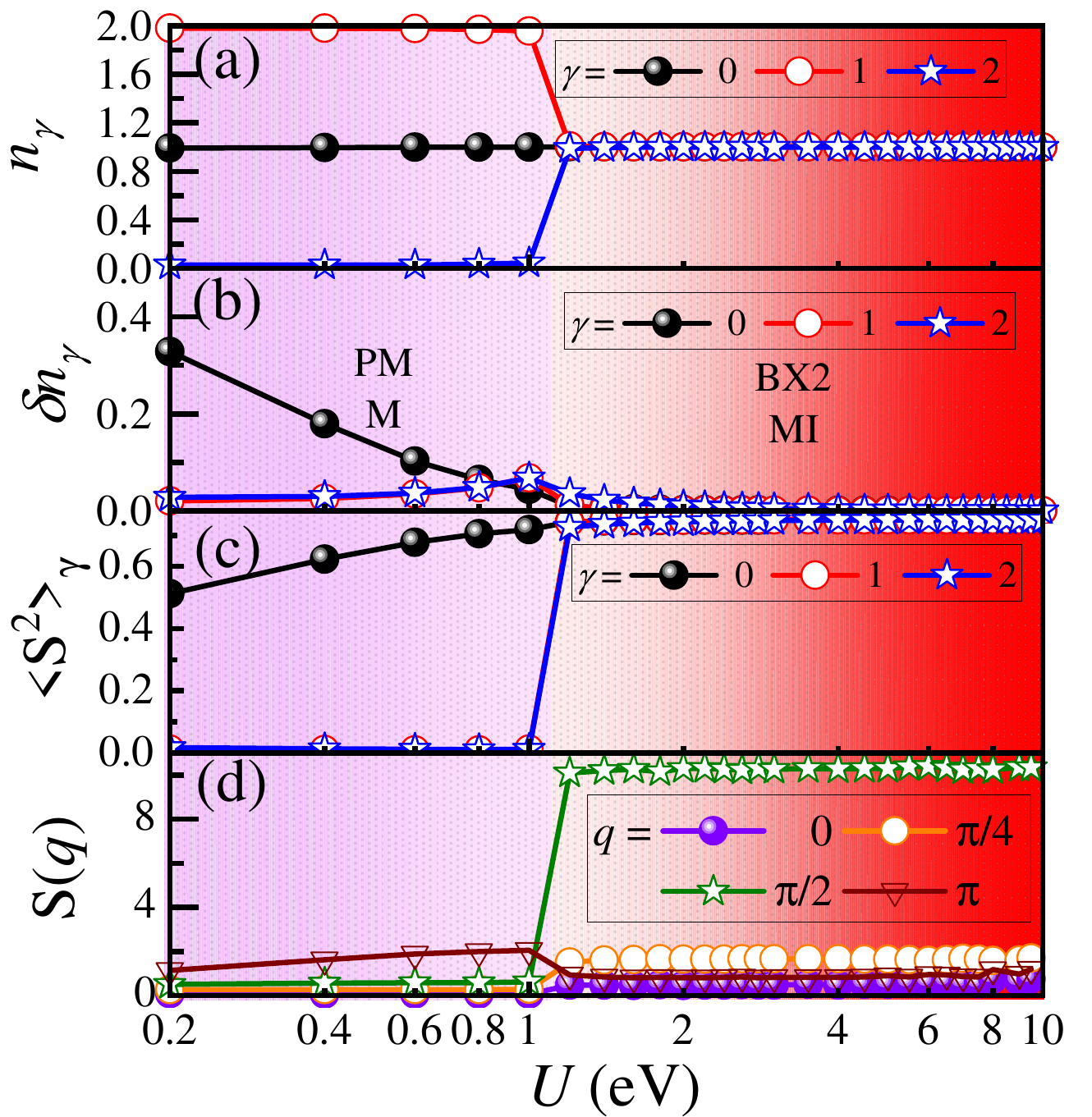}
\caption{(a) Orbital-resolved occupation number $n_{\gamma}$, (b) charge fluctuations $\delta{n_{\gamma}}=\frac{1}{L}\sum_{i}({\langle}n_{\gamma,i}^2\rangle-{\langle}n_{\gamma,i}{\rangle}^2)$, (c) $\langle S^2_{\gamma} \rangle$, and (d) spin structure factor $S(q)$ vs. $U$, at $J_{\rm H}/U = 0.2$. Here, we used a $16$-sites cluster chain with NN and NNN hoppings for three electrons in three orbitals.}
\label{n_u}
\end{figure}

We plot the site-average occupancy of different orbitals $n_{\gamma}$ for different values of $U$, as shown in Fig.~\ref{n_u}(a). At small $U$
($\textless 1$ eV), the $\gamma = 0$ ($d_{3z^2-r^2}$) orbital remains half-filled and the $\gamma = 1$ ($d_{yz}$) orbital is double occupied, while the $\gamma = 2$ ($d_{x^2-y^2}$) orbital is unoccupied (see Fig.~\ref{n_u}(a)). In this $U$ region, the spin correlation $S(r)$ decays rapidly as site distance $r$ increases, indicating paramagnetic behavior, while the charge fluctuations are mainly contributed by the $\gamma = 0$ ($d_{3z^2-r^2}$) orbitals (see Fig.~\ref{n_u}(b)).

By increasing the Hubbard interaction $U$, the population of all three orbitals reaches $1$ without charge fluctuations, as displayed in Fig.~\ref{n_u}(b), indicating a Mott insulating behavior. The strong local magnetic moments are fully developed with spin-squared $\langle S^2_{\gamma} \rangle$ =0.75 for each of the three orbitals when $U \textgreater 1$ eV, as shown in Figs.~\ref{n_u}(c).  In addition, we also summarize the spin structure factor $S(q)$ for different vectors as a function of $U$ in Fig.~\ref{n_u}(d).  In the small-$U$ paramagnetic phase ($U/W \textless 1$), all the $S(q)$s of different phases have similar values and do not display any obvious peak at a specific value of $q$, suggesting a PM state. When $U \textgreater 1$ eV, $S(\pi/2)$ becomes clearly dominant at the $U$ region we studied. Thus, this PM to Block transition is a metal to insulator transition, due to the absence of charge fluctuations in the latter, indicating this phase  should be a Block AFM Mott insulator.

Finally, let us briefly discuss the connection of our results with experimental results of the noncentrosymmetric polar materials Ba$M$Te$_2$O$_7$ ($M$ = Ni and Cu). Namely, using the same hoppings and crystals fields of our present
calculations, we can crudely estimate the properties of using other transition metals by simply
changing the electronic filling.
Following this procedure, in the $S = 1$ BaNiTe$_2$O$_7$ compound, the extra electron will occupy the lower $d_{yz}$ energy level, leading to only two ``active'' orbitals ($d_{3z^2-r^2}$ and $d_{xy}$). Thus, the AFM interaction strength of the NNN sites is still larger than that among the NN sites, leading to block coupling along the zigzag direction. For the $S = 1/2$ BaCuTe$_2$O$_7$ compound, now only one $d_{xy}$ orbital remains active. However, the hopping of $d_{xy}$ between NNN sites is about 40 times larger than the hopping of $d_{xy}$ between NN sites, leading to a quite weak magnetic coupling among the NN sites. Thus, this $S = 1/2$ zigzag chain may not form long-range magnetic order along the chain direction. We also would like to remark that the presence of additional interactions, such as interchain coupling, single-ion anisotropy or other effects, is necessary to stabilize long-range magnetic order and also important for the spin canting in the real bulk materials~\cite{Li:scpma,Chen:prm}. Otherwise in a one-dimensional system, the correlations always decay like a power law. These additions (single-ion anisotropy, etc.) are not the focus of the present work, thus we leave this issue to future studies.

\subsection{Additional DFT discussion}

\begin{figure}
\centering
\includegraphics[width=0.48\textwidth]{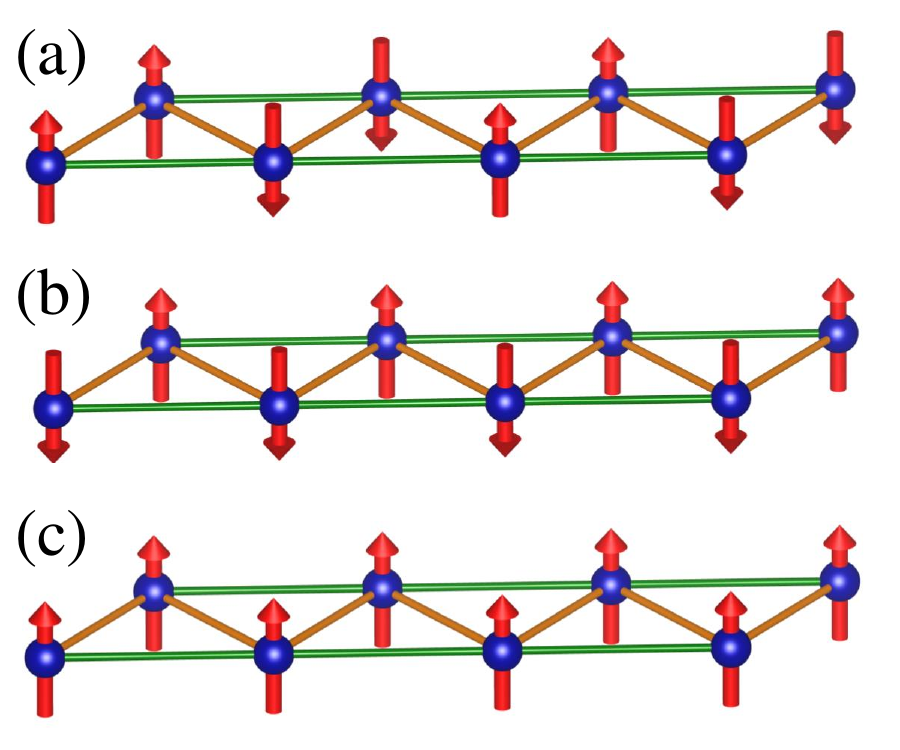}
\caption{Sketch of the three possible magnetic configurations (spins denoted by arrows) in the zigzag chain studied
via DFT+$U$: (a) Block AFM
with wavevector $k_q$ = $\pi/2$, (b) AFM1 with wavevector $k_q$ = $\pi$, and (c) FM with wavevector $k_q$ = 0.
Note that in (a) the pairs of spins pointing along the same direction (such as the pair pointing up on the far left)
could be located along the other diagonal of the zigzag chain as well, giving to this state a degeneracy two that
may lead to ``nematic'' consequences at finite temperature as it occurs for iron superconductors.}
\label{mag_fig}
\end{figure}

For completeness, let us briefly discuss our DFT magnetic results here. As shown in Fig.~\ref{mag_fig}, three possible magnetic configurations in the zigzag chain were considered: Block AFM with wavevector $k_q$ = $\pi/2$, AFM1 with wavevector $k_q$ = $\pi$, and FM with wavevector $k_q$ = 0.  In addition, the LSDA plus $U_{\rm eff}$ with the Dudarev format was introduced to simulate the on-site interactions, where $U_{\rm eff} = 6$ eV was used as discussed in the previous experimental work for BaCoTe$_2$O$_7$~\cite{Li:scpma}. Both the lattice constants and atomic positions were fully relaxed for those different spin states.

\begin{figure}
\centering
\includegraphics[width=0.48\textwidth]{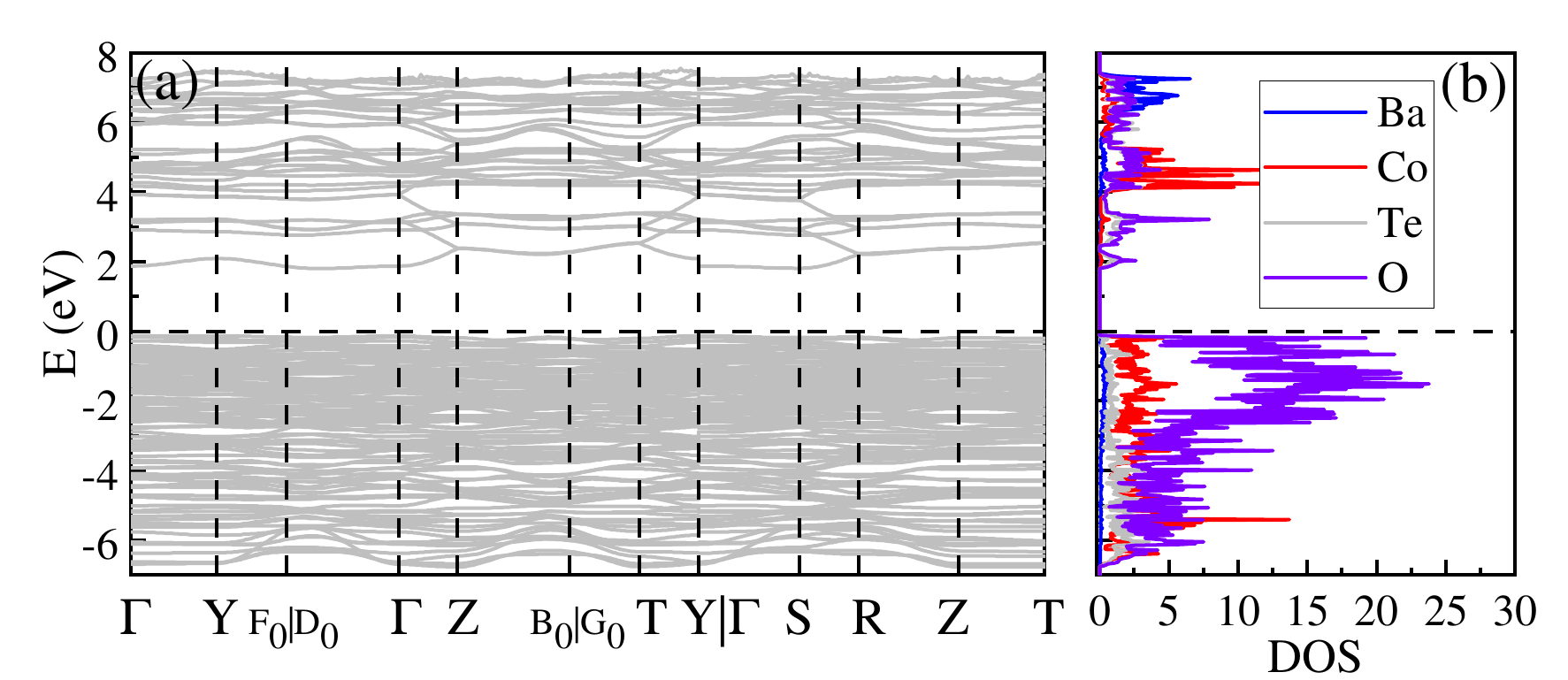}
\caption{(a) Band structure of BaCoTe$_2$O$_7$ for the Block AFM magnetic state with $U_{\rm eff} = 6$ eV. The Fermi level is shown with a dashed horizontal line. The coordinates of the high-symmetry points in the Brillouin zone are $\Gamma$ = (0, 0, 0), Y = (0.5, 0.5, 0), ${\rm F}_0$ = (0.30769, 0.69231, 0), ${\rm D}_0$ = (-0.30769, 0.30769, 0), Z = (0, 0, 0.5), ${\rm B}_0$ = (-0.30769, 0.30769, 0.5), ${\rm G}_0$ = (0.30769, 0.69231, 0.5), T = (0.5, 0.5, 0.5), S = (0, 0.5, 0), and R = (0, 0.5, 0.5). Note that all the high-symmetry points are in scaled units, corresponding to the units of $2\pi$/$s$, ($s = a, ~b$ or $c$). (b) Density-of-states near the Fermi level of BaCoTe$_2$O$_7$  for the Block AFM magnetic state with $U_{\rm eff} = 6$ eV. Note that only the spin up channel is displayed for both band structure and density of states here because the spin down channel is symmetric.}
\label{dosband_mag}
\end{figure}

First, the Block AFM magnetic order has the lowest energy, while the AFM1 and FM have a higher energy by about $\sim 15.5$ and $\sim 3.3$ meV/Co, respectively. Furthermore, we also calculated the local magnetic moment of Co atoms and it is $2.737$ $\mu_B$/Co, in reasonable
agreement with the  $S = 3/2$ spin state found in the model study. In addition, we also studied the band structures and density-of-states for the Block AFM magnetic state by using LSDA+$U$~\cite{Dudarev:prb} with $U_{\rm eff} = 6$ eV.  The calculated indirect band gap is $\sim 2$ eV, in good agreement with previous experimental studies using the UV-vis absorption spectrum that reported $\sim 2.68$ eV~\cite{Li:scpma}. These results support the  charge-transfer picture discussed in the previous section. Without any interaction, the Co's $3d$ states mainly contribute to the states near the Fermi level where most O'$2p$ states are located in a lower energy region with a large charge-transfer energy from O-$2p$ to Co-$3d$ orbitals. By introducing the Hubbard $U$, the Co-$3d$ states display Mott-insulating behavior with a large Mott gap ($\sim$ 8 eV at $U_{\rm eff} = 6$ eV), pushing the O-$2p$ states (slightly hybridized with Co-$3d$ states) close to the Fermi surface (see Fig.~\ref{dosband_mag}). Note that the spin dependence of the correlation energy density is already considered in the LSDA portion. As an overall effect, the calculated band gap of the system is only about 2 eV at a larger $U_{\rm eff}$ = 6 eV, much smaller than the Mott gap of the $3d$ states.

\section{Conclusions}
In summary, we systematically studied the zigzag compound BaCoTe$_2$O$_7$ by using first-principles DFT and DMRG calculations. Based on first-principles DFT, a strongly anisotropic one-dimensional electronic band structure was observed in the non-magnetic phase, corresponding to its dominant zigzag chain geometry. Furthermore, the $d_{xy}$ bands are more dispersive than other
orbitals' bands, suggesting that the $d_{xy}$ orbitals play the key role in magnetism and other physical properties in BaCoTe$_2$O$_7$. Based on the Wannier functions calculated from DFT, we obtained the NN and NNN hopping amplitudes and on-site energies for the cobalt atoms. The hopping of $d_{xy}$ to $d_{xy}$ between NNN Co-Co sites is the largest element in the hopping matrices, which is caused by the super-super
exchange via the path $d_{xy}$-$p_x$/$p_y$-$p_x$/$p_y$-$d_{xy}$.

Then, a multi-orbital Hubbard model for the cobalt chain was constructed and studied by using the many-body DMRG methodology, considering quantum fluctuations, for two models: (1) considering only a NN hopping matrix and (2)
considering NN plus NNN hopping matrices. Based on these DMRG calculations,  we obtained a robust staggered $\uparrow$-$\downarrow$-$\uparrow$-$\downarrow$ antiferromagnetic (AFM1) state when having only the NN hopping matrix in the chain direction, while a more dominant block (BX2) $\uparrow$-$\uparrow$-$\downarrow$-$\downarrow$ order was unveiled by introducing the NNN hopping matrix. At small Hubbard coupling strengths, this system displayed PM metallic phase behavior with large nonzero charge fluctuations contributed mainly by the $\gamma = 0$ ($d_{3z^2-r^2}$) orbital. At larger $U$, the system displays Mott insulator characteristics, due to the absence of charge fluctuations,
with three half-filled orbitals, in the region where the magnetic block AFM is obtained.

\section{Acknowledgments}
This work was supported by the U.S. Department of Energy (DOE), Office of Science, Basic Energy Sciences (BES), Materials Sciences and Engineering Division. G. A. was supported by the U.S. Department of Energy, Office of Science, National Quantum Information Science Research Centers, Quantum Science Center; he contributed his expertise with the DMRG algorithm, its applicability to the multi-orbital zig-zag chain, and the software implementation.

\section{APPENDIX}
\begin{figure}
\centering
\includegraphics[width=0.48\textwidth]{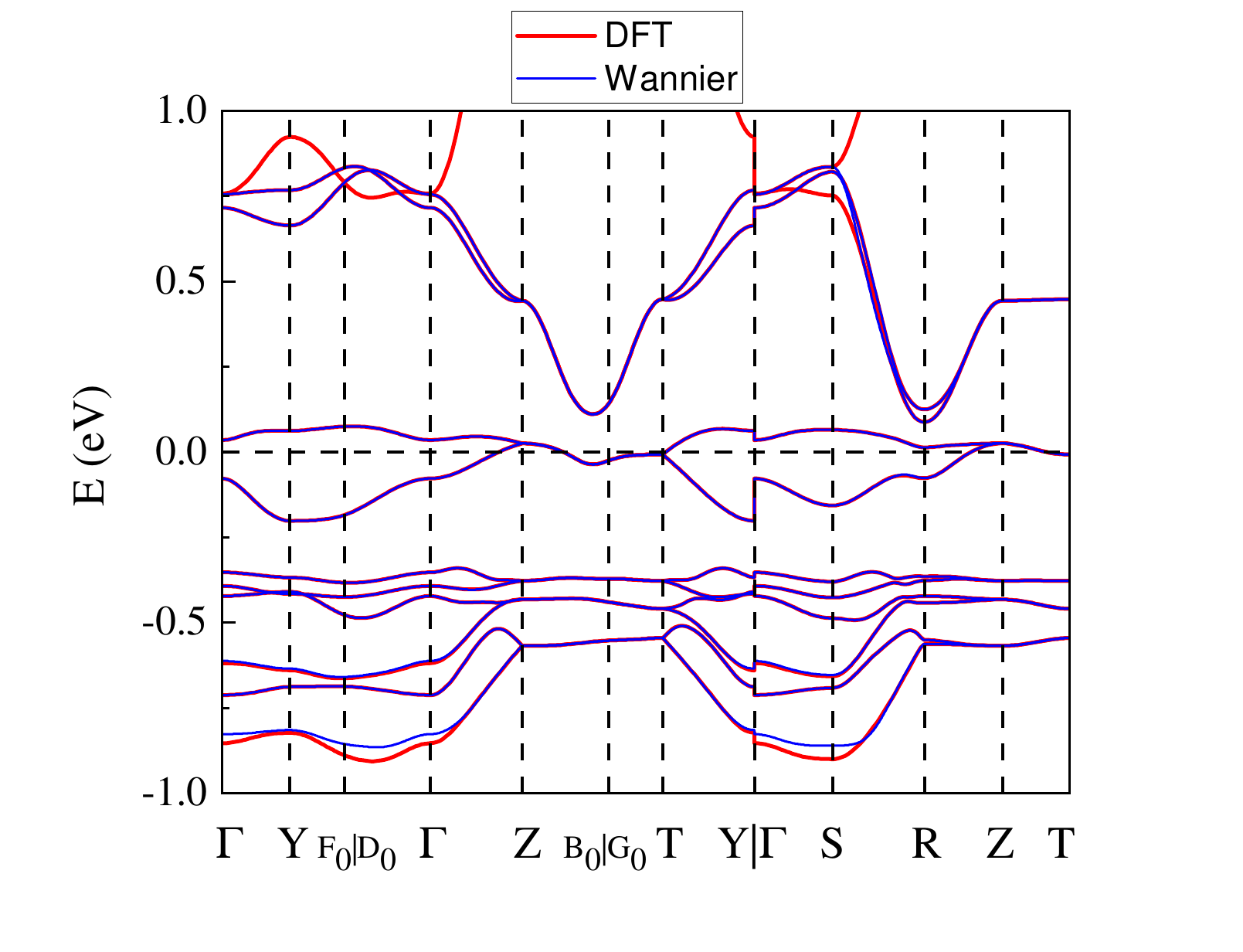}
\caption{(a) DFT and Wannier bands of BaCoTe$_2$O$_7$. The coordinates of the high-symmetry points in the Brillouin zone are  $\Gamma$ = (0, 0, 0), Y = (0.5, 0.5, 0), $F_0$ = (0.30769, 0.69231, 0), $D_0$ = (-0.30769, 0.30769, 0), Z = (0, 0, 0.5), $B_0$ = (-0.30769, 0.30769, 0.5), $G_0$ = (0.30769, 0.69231, 0.5), T = (0.5, 0.5, 0.5), S = (0, 0.5, 0), and R = (0, 0.5, 0.5). Note that all the high-symmetry points are in scaled units, corresponding to the units of $2\pi$/s, ($s = a, ~b$ or $c$). Note that the DFT electronic structures are calculated using the experimental crystal structure~\cite{Li:scpma} without and
additional Hubbard $U$.}
\label{wannier_fit}
\end{figure}

As shown in Fig.~\ref{wannier_fit}, the Wannier band structure can be fit well with the DFT bands of BaCoTe$_2$O$_7$. Based on the Wannier fitting results, we deduced the hopping parameters and on-site crystal fields. Here, the two largest hopping elements that we obtained are: $\sim 0.124$ eV between $d_{xy}$ orbitals for the NNN sites and $\sim 0.079$ eV between $d_{3z^2-r^2}$ orbitals for the NN sites, while other hopping elements are much smaller. Those two states ($d_{xy}$ and $d_{3z^2-r^2}$) are the key orbitals to understand this system. Due to similar crystal-splitting energies for $d_{yz}$ ($\sim -0.397$ eV), $d_{xz}$ ($\sim -0.527$ eV), and $d_{x^2-y^2}$ ($\sim -0.535$ eV), it is possible for the reordering of those orbitals in some $U$ and $J_{\rm H}$ regions. However, no matter which orbital is chosen ($d_{yz}$, $d_{xz}$ and $d_{x^2-y^2}$), it will not alter the calculational results, because of the nature of the hopping matrix. Note in our DMRG calculations, we considered a three-orbital with the basis ($d_{3z^2-r^2}$, $d_{yz}$, and $d_{xy}$) orbitals.

Here, we also list the hopping matrix with using the basis ($d_{3z^2-r^2}$, $d_{xz}$, $d_{yz}$, $d_{x^2-y^2}$, and $d_{xy}$) orbitals.
\begin{equation}
\begin{split}
t^{NN1}_{\gamma\gamma'} =
\begin{bmatrix}
    -0.079	&    -0.045	 &    0.027	  &  -0.043	  &   0.028	\\
     0.045	 &    0.044	 &   -0.023	  &  -0.010	   & -0.013	\\
     0.027	 &    0.023	 &    0.022	   & -0.005	   &  0.009	\\
    -0.043	 &    0.010	 &   -0.005	  &  -0.005	   & -0.001	\\
    -0.028	&    -0.013	 &   -0.009	  &   0.001	  &  -0.003	
\end{bmatrix}.\\
\end{split}
\end{equation}
\begin{equation}
\begin{split}
t^{NN2}_{\gamma\gamma'} =
\begin{bmatrix}
    -0.079	 &   0.045	 &   -0.027	    &-0.043	 &   0.028	\\
    -0.045	 &    0.044	  &  -0.023	   & 0.010	 &    0.013	\\
    -0.027	 &    0.023	  &   0.022	   &  0.005	  &   -0.009	\\
    -0.043	 &    -0.010 &  0.005	   & -0.005	  &   -0.001	\\
    -0.028	  &   0.013	   & 0.009	   & 0.001	  &  -0.003	
\end{bmatrix}.\\
\end{split}
\end{equation}
\begin{equation}
\begin{split}
t^{NNN}_{\gamma\gamma'} =
\begin{bmatrix}
    -0.026	&     0.015	  &  -0.007	    & 0.007	   &  0.019	\\
     0.015	&    -0.022	  &   0.036	    & 0.007	    &-0.042	\\
     0.007	&    -0.036	  &   0.013	    &0.012	   & -0.038	\\
     0.007	&     0.007	  &  -0.012	   & -0.060	   &  0.004	\\
    -0.019	&     0.042	  &  -0.038	   & -0.004	    & 0.124	
\end{bmatrix}.\\
\end{split}
\end{equation}

\end{document}